\documentclass{SciPost}

\hypersetup{
    colorlinks,
    linkcolor={red!50!black},
    citecolor={blue!50!black},
    urlcolor={blue!80!black}
}

\usepackage[bitstream-charter]{mathdesign}
\DeclareSymbolFont{usualmathcal}{OMS}{cmsy}{m}{n}
\DeclareSymbolFontAlphabet{\mathcal}{usualmathcal}

\fancypagestyle{SPstyle}{
\fancyhf{}
\lhead{\colorbox{scipostblue}{\bf \color{white} ~SciPost Physics }}
\rhead{{\bf \color{scipostdeepblue} ~Submission }}

\fancyfoot[C]{\textbf{\thepage}}
}

\usepackage[
activate={true,compatibility},
final,
tracking=true,
kerning=true,
nopatch=footnote]{microtype}

\usepackage{mathtools}
\usepackage{amsfonts}
\usepackage{nicefrac}
\usepackage{physics}
\usepackage{color}
\usepackage{graphicx}
\usepackage{dcolumn}

\usepackage[ruled,vlined]{algorithm2e}
\usepackage{algpseudocode}
\usepackage{hyperref}

\usepackage{tabularx}
\usepackage{booktabs}

\newcolumntype{M}{>{\centering}X}
\newcolumntype{Y}{>{\hsize=.2\linewidth\arraybackslash}X}
\newcolumntype{C}{>{\hsize=.24\linewidth\centering\arraybackslash}X}
\newcolumntype{R}{>{\hsize=.125\textwidth\raggedleft\arraybackslash}X}

\usepackage{subcaption}
\begin{document}

\pagestyle{SPstyle}

\begin{center}{\Large \textbf{\color{scipostdeepblue}{
Quantum-inspired space-time PDE solver and dynamic mode decomposition\\
}}}\end{center}

\begin{center}\textbf{
Raghavendra D. Peddinti\textsuperscript{1,$\star$},
Stefano Pisoni\textsuperscript{1,2},
Narsimha Rapaka\textsuperscript{3},
Yacine Addad\textsuperscript{3},
Mohamed K. Riahi\textsuperscript{3},
Egor Tiunov\textsuperscript{1} and
Leandro Aolita\textsuperscript{1}
}\end{center}

\begin{center}
{\bf 1} Quantum Research Center, Technology Innovation Institute, Abu Dhabi, UAE
\\
{\bf 2} Hamburg University of Technology, Institute for Quantum-Inspired and Quantum Optimization, Germany
\\
{\bf 3} Emirates Nuclear Technology Center, Khalifa University of Science and Technology, Abu Dhabi, UAE
\\[\baselineskip]
$\star$ \href{mailto:raghavendra.peddinti@tii.ae}{\small raghavendra.peddinti@tii.ae}\,
\end{center}

\section*{\color{scipostdeepblue}{Abstract}}
\textbf{\boldmath{%
The numerical solution of time-dependent PDEs posed over the combined space-time domain, instead of treating time sequentially, leads to space-time methods. Treating time as another spatial dimension allows parallelization across time and simultaneous adaptivity in spatial and temporal discretizations.  However, space-time formulations suffer from the \textit{curse of dimensionality} due to the (d+1)-dimensional domain. Time-series forecasting methods, such as dynamic mode decomposition (DMD), also suffer from the same scaling for storage and processing of measurement data.
In this work, we investigate whether quantum-inspired tensor network methods can mitigate this curse for both solving PDEs and data-driven modeling.
We formulate a space-time matrix product state (MPS) encoding to represent either the space-time solution or time-series data. Using this representation, we develop a unified framework that combines an MPS space-time PDE solver with an MPS-based dynamic mode decomposition (MPS-DMD) algorithm.
We benchmark our MPS space-time solver for linear and nonlinear PDEs in both (1+1) and (2+1) dimensions, including non-diffusive regimes such as shockwave formation and Allen-Cahn phase separation. We observe that the MPS ansatz efficiently captures the spatio-temporal correlations of the solutions, with a  $99\%$ reduction in parameters with respect to the grid size.
We demonstrate the MPS-DMD algorithm in two cases. First, the MPS space-time solver generates the MPS encoding of the time-series data. We directly apply MPS-DMD to predict the dynamics without solving PDEs, showcasing the complementary nature of both methods. Second, for the general case where data is available as dense arrays, MPS-DMD needs an intermediate compression step. We implement a streaming-like compression algorithm and observe a $97\%$ reduction in memory, with MPS-DMD predictions matching those of standard DMD. 
}}

\vspace{\baselineskip}

\vspace{10pt}
\noindent\rule{\textwidth}{1pt}
\tableofcontents
\noindent\rule{\textwidth}{1pt}
\vspace{10pt}

\section{Introduction}

The evolution of dynamical systems is described by time-dependent partial differential equations (PDEs) whose exact solution is seldom available, and numerical methods provide approximations. Numerical solvers often discretize time separately from spatial dimensions, resulting in time-stepping schemes~\cite{iserles2009first}. 
In contrast, space-time methods treat time as an additional dimension of the computational domain and represent the solution in (d+1)-dimensions~\cite{SchafelnerAndreas2021SFEM}. Space-time methods offer greater scope for parallelization and enable local adaptivity in space and time discretizations~\cite{wick2023space, SchafelnerAndreas2021SFEM}. Despite these benefits, the curse of dimensionality hinders the practical applications of space-time methods. This curse is also manifested in data-driven methods, which provide an alternative tool when the mathematical models (PDEs) are either unavailable or not accurate.

Quantum-inspired tensor network methods can mitigate this curse of dimensionality. They were originally developed in quantum many-body physics to avoid the exponential growth of the Hilbert space~\cite{orus2014practical}. Tensor networks have since expanded into a powerful mathematical framework~\cite{cichocki2016tensor, cichocki2017tensor}. The most successful tensor network is a 1D chain of tensors known as matrix product states (MPSs) or tensor trains (TTs)~\cite{schollwock2011density,verstraete2008matrix,oseledets2011tensortrain}.
The field of MPS-based PDE solvers is growing rapidly~\cite{kornev2023chemicalmixer, peddinti, gourianov2022quantum, kiffner_jaksch2023tensor, Erika_vlasov_eq, van2025qinsp, truong2023tensor, van2026quantum}. The central idea is to represent discretized functions and operators using the QTT encoding~\cite{garcia_ripoll2021quantum}. QTTs are a variant of MPS for representing vectors/matrices resulting from the discretization of functions/operators on a uniform mesh. QTTs compress the solution by capturing correlations across the spatial scales~\cite{Image_Compression,gourianov2022_thesis}. 
The QTT solution is evolved within this compressed representation using time-stepping schemes. The timestep ($\Delta t$) of explicit schemes scales with spatial discretization ($\Delta x$) due to the Courant-Friedrichs-Lewy (CFL) condition~\cite{cfl_1928}. Hence, the high spatial resolution afforded by QTT compression is lost in the large number of timesteps necessary to reach the final duration~\cite{pisoni2025compression}. Implicit schemes enable larger timesteps but require a linear solve per time-step~\cite{oseledets2012solution,  rodgers2023implicit}. Alternatively, space-time methods provide the solution for all timesteps at once using a global linear solve. This motivates the application of MPSs as an efficient ansatz for space-time solutions~\cite{Dolgov_st2012, Breiten2021, adak2024tensor, Adak_2025, arenstein2025fast}. However, the development of MPS-based space-time solvers is currently in its nascent phase. It is unclear whether MPSs can encode solutions from diverse PDE regimes.   

On the other hand, dynamic mode decomposition (DMD) is a data-driven tool for time-series forecasting~\cite{SCHMID_DMD_2010, kutz2016dynamic}. The DMD algorithm determines the dynamically relevant spatial modes and uses these modes to predict future evolution. These modes also contain diagnostic information about the underlying dynamics. The time-series data is organized into snapshot matrices, which scale linearly with the spatial and temporal resolution. Tensor networks provide an efficient ansatz for these snapshot matrices~\cite{klus2018tensor}. However, the data must still be converted into the MPS encoding to reduce DMD's memory and runtime requirements.

\begin{figure}
    \centering
    \includegraphics[width=0.7\linewidth]{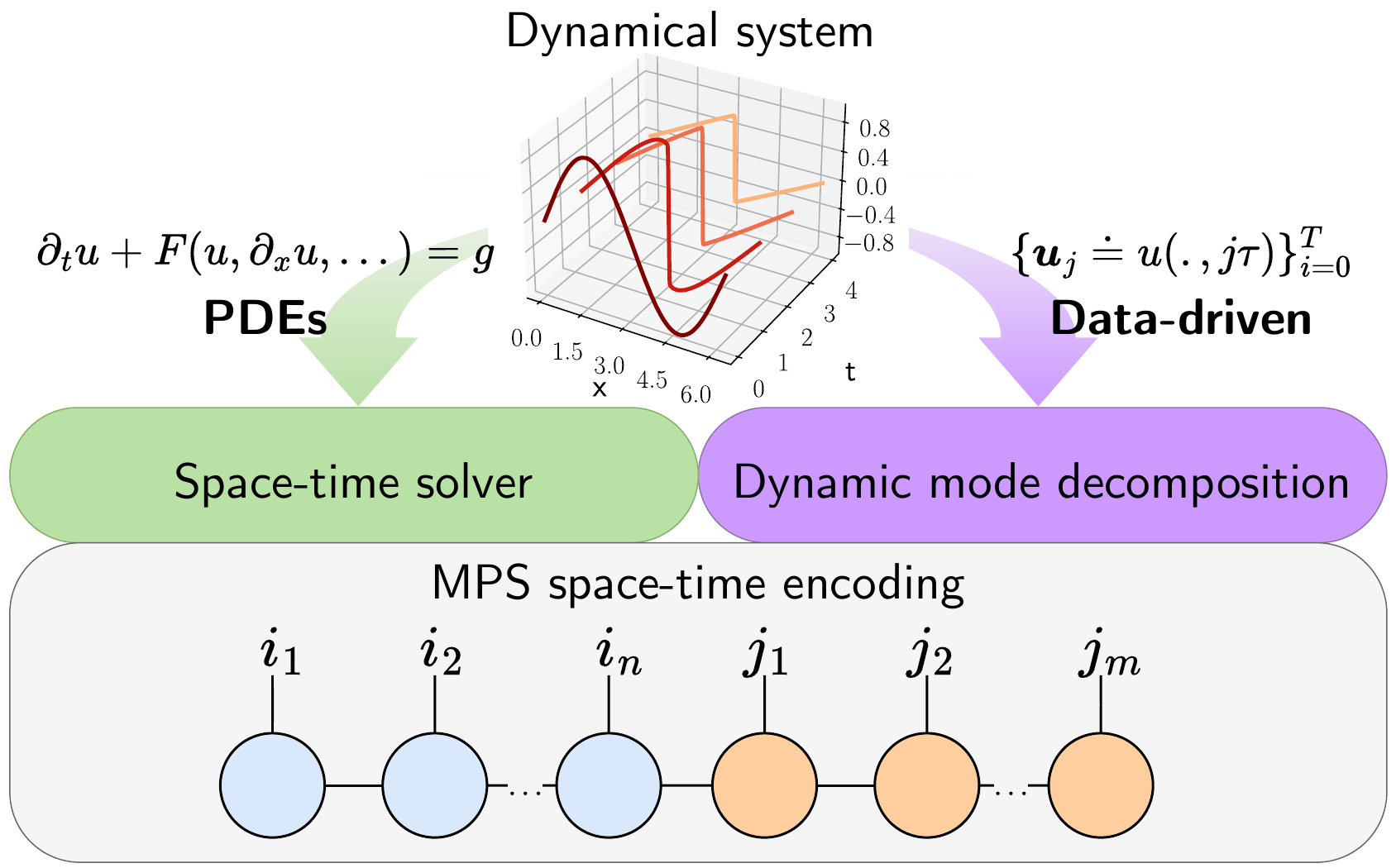}
    \caption{\textbf{Overview of simulation and prediction of dynamical systems, using the MPS space-time encoding.}
    Let $u(x,t)$ denote the state of the dynamical system. The evolution of the state can be simulated by solving the governing equations, given by time-dependent PDEs. Alternatively, the dynamics can be predicted by learning from the system's historical data, represented by a time series. Using matrix product state (MPS) encoding over the combined space-time domain, we develop MPS counterparts to the standard space-time solver and dynamic mode decomposition.
    }
    \label{fig:schematic}
\end{figure}

In this work, we unify MPS-based space-time methods for both simulation and data-driven prediction. In Fig.~\ref{fig:schematic}, we present a schematic of our contributions. First, we introduce an MPS space-time encoding that can represent either the space-time solution or the time-series data. Using this representation as the building block, we develop two methods: a) an MPS space-time solver and b) an MPS-DMD algorithm for solving PDEs and DMD predictions, respectively.
We develop the MPS space-time solver for both linear and nonlinear PDEs. We benchmark our solver for (1+1) and (2+1) dimensional PDEs and consider a diverse set of dynamical regimes from classical physics: linear and nonlinear diffusion, wave transport, phase separation, and shockwave formation. We also consider the nonlinear Schrödinger equation from quantum dynamics. For the above PDEs, we observe that the MPS solution can capture most of the relevant dynamics using small bond dimensions, compressing more than $99\%$ of the original space-time grid size. We also study the spatio-temporal correlations of the solution, allowing us to interpret the compression power of MPSs.

We reformulate the DMD algorithm entirely within the MPS space-time representation. This allows us to develop an MPS-DMD algorithm in two different settings. First, using the MPS space-time solver, we efficiently generate the time-series data necessary for the DMD method. This combination is particularly beneficial for systems whose measurement data is unavailable, and we demonstrate this by predicting the long-term dynamics of the 1D Burgers' equation. Second, we consider the case of Karman vortex street simulation whose time-series data is available as dense vectors~\cite{dataset_cylinder}. We compress this data into a space-time MPS using a streaming-like algorithm and apply MPS-DMD, leading to a $97.42\%$ reduction in memory usage. Under the assumption of low spatial and temporal entanglement, our findings imply exponential memory and runtime advantages over standard PDE solvers and DMD methods.

\paragraph{Connection with previous works:}We summarize the earlier works on tensor network-based space-time solvers, and comment on the connection with our work. \textit{Dolgov et al.}~\cite{Dolgov_st2012} presented a global space-time TT formulation for the Fokker-Planck equation. \textit{Breiten et al.}~\cite{Breiten2021} implemented a TT space-time method for solving nonlinear control problems. \textit{Adak et al.}~\cite{adak2024tensor, Adak_2025} demonstrated the advantages of TT space-time spectral solvers for both linear and nonlinear transport problems. Recently, \textit{Arenstein et al.}~\cite{arenstein2025fast} studied the 1D Burgers’ equation, demonstrating clear advantages over standard time-stepping schemes. In this work, we study a wide range of linear and nonlinear PDEs and independently observe similar advantages. In particular, we use the QTT representation whose memory scales logarithmically with the original grid size and linearly with dimension for a fixed bond dimension. Additionally, none of these works connect the MPS-DMD algorithm with the space-time solution. Although \textit{Klus et al.}~\cite{klus2018tensor} developed a tensor-based DMD algorithm, it leaves the temporal dimension uncompressed, leading to a linear complexity in time. In a preprint that appeared online during this work, \textit{\'Sroda et. al.}~\cite{sroda2025predictor} independently developed the MPS-DMD algorithm.

\section{Preliminaries}
\subsection{Space-time methods}\label{sec:space-time-methods}
We introduce the standard space-time formulation for solving a (1+1)-dimensional time-dependent PDE of $u$ defined on the domain $(x, t) \in \Omega \times(0,T]$, given by:
\begin{equation}\label{eq:general_pde}
    \partial_t^p u + F(u,\partial_x u,\dots,\partial_x^{q} u) = g(x,t)
\end{equation}
where $F$ is a function of the solution $u$ and $\{\partial_x^{k}\}_{k=1}^q$, its partial derivatives in space up to order $q$. $\partial^p_t$ refers to the $p$-th derivative in time (usually, $p=1$ or $2$), and $g(x,t)$ defines the source term. The corresponding initial and boundary conditions are given by $u_0(x)$ (and $\partial_t u(x,t)|_{t=0}$ for the second derivative in time) and $u(x,t)|_{\partial \Omega}$, respectively.
After discretization, the objective is to find the solution simultaneously for all $N_x \times N_t$ points in the discretized domain $\{h, ..., N_x h\} \times \{\tau, ..., N_t \tau\}$, where $h = |\Omega| / N_x$ and $\tau = T / N_t$. 

First, we use a finite difference scheme to compute the time derivative with forward differences and the spatial derivatives with central differences.
Next, we build the \textit{all-at-once} system by assembling the discretized space and time operators of Eq.~\eqref{eq:general_pde} using the tensor product ($\otimes$). The resulting system, along with the appropriate initial and boundary condition terms $\mathcal{I}(u),\mathcal{B}(u)$, is given by:
\begin{equation}\label{eq:all_at_once_linsys}
    \left(\mathbb{I}_x \otimes \mathbb{D}_t^p + \mathbb{F}_x(u) \otimes \mathbb{I}_t \right) \boldsymbol{u} = \boldsymbol{g} + \mathcal{I}(u)+\mathcal{B}(u),
\end{equation}
where $\mathbb{I}$ is the identity operator, $\mathbb{D}^p_t$ is the discretized time derivative of order $p$, and $\mathbb{F}_x(u)$ is the discretized spatial operator as a function of $u$.
Here, $\boldsymbol{u}$ and $\boldsymbol{g}$ denote the discretized versions of $u(x,t)$ and $g(x,t)$, which are vectorized as:
\begin{equation} \label{eq:explicit_solution}
\boldsymbol{u} = (u_{1,1}, u_{1,2}, ..., u_{1,N_t}, u_{2,1}, ..., u_{N_x,N_t})^T     
\end{equation}
where, $u_{i,j}=u(ih, j\tau)$ and analogously for $\boldsymbol{g}$.
Moreover, Eq.~\eqref{eq:all_at_once_linsys} can be solved with a linear system solver only if the equation is linear, i.e., $\mathbb{F}_x$ is independent of $u$.
For nonlinear PDEs, the algebraic system is linearized according to the Picard iterative scheme~\cite{Golub1992}. This is implemented by replacing the dependence of $\mathbb{F}_x(u)$ by the solution at previous iteration, i.e., $\mathbb{F}_x (u^{(p-1)})$ at iteration $p$. The initial guess is chosen by repeating the initial condition for all time steps: $u^{(0)} = u_{0} \otimes I_t$.

Throughout our work, we employ the implicit formulation of the problem. In Appendix~\ref{app:space-time}, we present the detailed construction of the all-at-once linear system for both linear and nonlinear equations. Additionally, we note that the generalization to higher dimensions follows a similar construction.

\subsection{Dynamic mode decomposition}\label{sec:dmd}
Dynamic mode decomposition (DMD) is a data-driven technique for analyzing nonlinear systems by revealing their underlying spatiotemporal characteristics through spectral analysis. Initially developed within the fluid dynamics community~\cite{SCHMID_DMD_2010}, the DMD toolbox has since expanded and found applications in various fields, including robotics, quantum physics, neuroscience, and data processing~\cite{Rowley2009, Schmid_2022,ilia_dmd}.
The success of DMD is theoretically underpinned by its connection with Koopman operators and their spectra. We recommend that the reader follow these outstanding reviews~\cite{colbrook2024multiverse,kutz2016dynamic} for a detailed understanding.

\begin{algorithm}
\begin{algorithmic}[1]
\Require Snapshot matrices

$X = [\boldsymbol{u}_0, \dots, \boldsymbol{u}_{N_t-1}] $, 
$X' = [\boldsymbol{u}_1, \dots, \boldsymbol{u}_{N_t}]$;~~~$X,X'\in\mathbb{R}^{N_x\times N_t}$
\Ensure $X'=AX \implies A:=X'X^{+}$ where $X^{+}$ is the pseudoinverse of $X$.
\State Proper orthogonal decomposition: $X = U \Sigma V^{\dagger}$, with rank $|\Sigma|$
\State Project $A$ onto the span of $U$: $\tilde{A} = U^{\dagger} A U= U^{\dagger} X' V \Sigma^{-1} \in \mathbb{R}^{|\Sigma|\times |\Sigma|}$
\State Eigen decomposition: $\tilde{A} W = W \Lambda$
\State Compute the dynamic modes: $\Phi = UW \in \mathbb{C}^{N_x\times|\Sigma|}$

\Return $(\Lambda, \Phi)$
\end{algorithmic}
\caption{DMD}\label{alg:exact-dmd}
\end{algorithm}
In Algorithm~\ref{alg:exact-dmd}, we summarize 
the standard DMD algorithm. This assumes two discrete time series of  states $\boldsymbol{u}_j\doteq u(\cdot,j\tau)\in \mathbb{R}^{N_x}$, one starting at time $j=0$ and the other one at time $j=1$, arranged respectively into $X$ and $X'$ in $\mathbb{R}^{N_x\times N_t}$, known as \emph{snapshot matrices}.
The (possibly nonlinear) dynamics between $X$ and $X'$ are approximated by a linear evolution operator $A$. 
First, the singular value decomposition, also known as \emph{proper orthogonal decomposition} (POD), of $X$ is computed. The singular values $\Sigma$ are then used to compute the pseudoinverse $X^+$, and $A:=X'X^+$ is given by $X'V^{\dagger}\Sigma^{-1}U$. By projecting $A$ onto the span of the left singular vectors $U$, one obtains a $|\Sigma|\times|\Sigma|$ matrix $\tilde{A}$. By computing the eigen decomposition of $\tilde{A}$, the \emph{dynamic modes} $\Phi$ of the data are found. These modes are then used to predict the future evolution of the system without having to solve any PDE. 
Moreover, in practice, $\Sigma$ is usually truncated to keep the $r$ most significant singular values, resulting in $r$ dynamic modes. This leads to an efficient approximation of $A$ that is often very accurate, even for $r \ll |\Sigma|$.

\section{Results}
As shown in Fig.~\ref{fig:schematic}, we develop two methods for simulating and predicting dynamics: the MPS space-time solver and MPS-DMD, respectively.
We first introduce the MPS space-time encoding. We then present the details of our MPS space-time solver, along with the numerical simulations of both linear and nonlinear PDEs. Finally, we introduce the MPS-DMD algorithm and predict long-term dynamics of nonlinear systems such as the Karman vortex street.

\subsection{MPS space-time encoding}
We briefly introduce the MPS space-time representation for encoding $M$, a 2D array of size $N_x\times N_t$.
It is assumed, without loss of generality, that $N_x$ and $N_t$ are powers of two, i.e., $N_x=2^n; N_t=2^m$.
Next, we binarize the indices of $M$, given by $i:=\{i_1,...,i_n\}$ and $j:=\{j_1,...,j_m\}$, assuming the big-endian form.
Then, $M$ is decomposed such that each element $M_{i,j}$ is given by:
\begin{equation}\label{eq:st_mps}
    M_{i,j} = A_1^{(i_1)}A_2^{(i_2)}\dots A_n^{(i_n)} A_{n+1}^{(j_1)}A_{n+2}^{(j_2)}\dots A_{n+m}^{(j_m)},
\end{equation}
where $A_l^{(0)}$ and $A_l^{(1)}$ constitute the $2(n+m)$ matrices of the MPS, with sizes ${\chi_{l-1}\times\chi_{l}}$. In our case, $\chi_0 =\chi_{n+m}=1$. 
We note that the ordering of the binary indices is in fact a critical choice, and different orderings lead to varying degrees of compression and convergence of the MPS solver. Unless specified otherwise, we follow the \textit{space-time ordering}, shown in Eq.~\eqref{eq:st_mps} and Fig.~\ref{fig:schematic}. This choice corresponds to Eq.~\eqref{eq:explicit_solution} in the vectorized form.

\subsection{MPS space-time solver}\label{sec:mps_spacetime}
Given an (1+1)D PDE with $p^\mathrm{th}$ derivative in time (typically, $p=1,2$), up to $q^\mathrm{th}$ spatial derivative, and source term $g(x,t)$:
\begin{equation}
    \partial_t^p u + F(u,\partial_x u,\dots,\partial_x^{q} u) = g(x,t),
\end{equation}
We develop an MPS solver following the outline of standard space-time methods~\cite{SchafelnerAndreas2021SFEM,wick2023space}. We refer to Section~\ref{sec:space-time-methods} for a brief overview of the space-time formulation.
The space-time domain is discretized by a uniform mesh. The discretized functions are encoded into the corresponding space-time MPSs, and the finite difference operators into matrix product operators (MPOs). The boundary and initial conditions are also encoded directly as MPS/MPOs. All the MPSs and MPOs are either constructed analytically~\cite{kazeev2012low, Kazeev2013} or found numerically using TT-cross interpolation~\cite{Savostyanov2011FastAI,ritter2024quantics}. SVD rounding is performed after each algebraic operation, capping the bond dimension at $\chi_{\mathrm{max}}$~\cite{oseledets2011tensortrain}.
For nonlinear PDEs, we use the Picard/Newton iterative update to linearize the equations~\cite{Golub1992}. Finally, we assemble the all-at-once linear system and solve it using DMRG-inspired methods~\cite{oseledets2012solution,dolgov2014alternating}. 
Additionally, in Appendix~\ref{app:space-time}, we present a detailed construction of the MPS space-time solver.

\begin{table}[t!]
\centering
\small
\begin{tabularx}{\textwidth}{XCCCC}
\toprule
PDEs & $\chi_{\max}$ & Comp. (\%) & $L_2$-error & Residual \\
\toprule
Heat: $u_t-\alpha u_{xx}=0$ & 2 & 99.98 & $2.1\times10^{-4}$ & $2.6\times10^{-12}$ \\
\multicolumn{5}{l}{\footnotesize$\alpha=1$, $t\in[0,1]$, $x\in[0,2\pi]$}\\
\multicolumn{5}{l}{\footnotesize$u_0(x)=\sin x$; $u(0,t)=u(2\pi,t)=0$} \\
\midrule
Wave: $u_{tt}-c^2u_{xx}=0$ & 2 & 99.98 & $5.3\times10^{-4}$ & $9.0\times10^{-8}$ \\
\multicolumn{5}{l}{\footnotesize$c=1$, $t\in[0,5]$, $x\in[0,10]$}\\
\multicolumn{5}{l}{\footnotesize$u_0(x)=\sin\pi x$; $u(0,t)=u(10,t)=0$} \\
\midrule
Burgers': $u_t+uu_x-\nu u_{xx}=0$ & 14 & 99.89 & $5.7\times10^{-4}$ & $5.1\times10^{-6}$ \\
\multicolumn{5}{l}{\footnotesize$\nu=10^{-3}$, $t\in[0,10]$, $x\in[0,1]$}\\
\multicolumn{5}{l}{\footnotesize$u_0(x)=\frac{2\nu\pi\sin(\pi x)}{1.01+\cos\pi x}$;  $u(0,t)=u(1,t)=0$} \\
\midrule
Burgers' (shockwave) & 20 & 99.85 & $4.0\times10^{-3}$ & $4.8\times10^{-4}$ \\
\multicolumn{5}{l}{\footnotesize$\nu=5\times10^{-3}$, $t\in[0,8.192]$, $x\in[0,2\pi]$}\\
\multicolumn{5}{l}{\footnotesize$u_0(x)=\sin x$; $u(0,t)=u(2\pi,t)=0$} \\
\midrule
Nonlinear diffusion: $u_t-(\kappa(u)u_x)_x=0$ & 20 & 99.85 & $3.0\times10^{-4}$ & $6.9\times10^{-6}$ \\
\multicolumn{5}{l}{\footnotesize$\kappa(u)=0.1e^{0.5u}$, $t\in[0,15]$, $x\in[1,3]$} \\
\multicolumn{5}{l}{\footnotesize$u_0(x)=\tfrac12(5-x)+(x-1)(x-3)$; $u(1,t)=2$, $u(3,t)=1$} \\ \addlinespace[3pt]

\midrule
Allen-Cahn: $u_t-\epsilon^2u_{xx}-u+u^3=0$ & 17 & 99.87 & $2.7\times10^{-5}$ & $3.1\times10^{-6}$ \\
\multicolumn{5}{l}{\footnotesize$\epsilon=0.05$, $t\in[0,0.1]$, $x\in[-1,1]$}\\
\multicolumn{5}{l}{\footnotesize$u_0(x)=e^{-x^2/0.18}$; BC: periodic} \\
\midrule
Allen-Cahn (spinodal IC) & 32 & 99.76 & $3.6\times10^{-4}$ & $2.5\times10^{-5}$ \\
\multicolumn{5}{l}{\footnotesize$\epsilon=0.15$, $t\in[0,4]$, $x\in[-1,1]$}\\
\multicolumn{5}{l}{\footnotesize$u_0(x)=\sum_{k=1}^{10}(a_k\sin\pi kx+b_k\cos\pi kx)/k$, $a_k,b_k\sim\mathcal{N}(0,0.4)$, BC: periodic} \\
\midrule
NLSE: $i\psi_t+\psi_{xx}+2|\psi|^2\psi=0$ & 20 & 99.85 & $3.0\times10^{-4}$ & $1.8\times10^{-2}$ \\
\multicolumn{5}{l}{\footnotesize$t\in[0,3]$, $x\in[-3\pi,3\pi]$}\\
\multicolumn{5}{l}{\footnotesize$\psi_0(x)=\operatorname{sech}(x+2)\,e^{i(x+2)/2}$; BC: periodic} \\
\bottomrule
\end{tabularx}
\caption{\textbf{Summary of PDEs solved using the space-time MPS solver.}  We solve each of the above PDEs with a grid size of $1024\times1024$, i.e., using a space-time MPS of $L_x+L_t=20$. All cases are run for 16 Picard/Newton iterations and 4 DMRG sweeps per iteration. We report the largest bond dimension ($\chi_{\max}$) of the converged solution, the MPS compression as the percentage reduction from the dense $N_x\times N_t$ array, along with the $L_2$-error to a reference solution and the final residual $\|Ax-b\|/\|b\|$ of the MPS solver. Below each PDE, we note the parameters, space and time domains, initial (IC) and boundary (BC) conditions.
The $L_2$-error is relative to the exact analytical solution when available or to a high-resolution numerical solution obtained using a standard time-stepping scheme.
For Burgers' and Allen-Cahn equations, we solve additional non-diffusive regimes that are challenging due to shockwave formation and phase separation, respectively. We refer to Appendix~\ref{app:benchmarks} for a detailed description of each benchmark.
}
\label{tab:equations}
\end{table}

\paragraph{Numerical results:}
In Table~\ref{tab:equations}, we present a summary of numerical results using the MPS space-time solver for both linear and nonlinear PDEs. For all PDEs, we use a $1024\times1024$ uniform grid, resulting in a 20-qubit MPS for the solution. The MPS space-time solver is run for 4 DMRG sweeps to solve the all-at-once linear system. For nonlinear equations, we use 16 Picard/Newton iterations. We note that the MPS solver achieves over 99\% memory compression compared to the dense vector space-time solver. In Appendix~\ref{app:benchmarks}, we present the detailed benchmarks for each instance of Table~\ref{tab:equations}. We compare the MPS space-time solution with analytical solutions whenever available. When analytical solutions do not exist, we use standard numerical solvers as our reference. Additionally, in App.~\ref{app:benchmarks_2d}, we extend our MPS solver to (2+1)D PDEs. In particular, we solve the 2D counterparts of the heat, Burgers', and Allen-Cahn equations and observe that MPSs are efficient even in higher dimensions. 

\paragraph{Bond dimension and compressibility:} 
To estimate the compressibility, we compute the bipartite entanglement entropy $S_{A|B}$ of the MPS space-time solution given by $S_{A|B}= -\sum_{j=1}^{\chi} s_j^2 \log_2 s_j^2$, where $s_j $ are the Schmidt coefficients of the MPS with the physical indices partitioned into $A$ and $B$ (See Ref.~\cite{orus2014practical} for more details).
We note that the entanglement entropy $S$ is upper bounded by $\log(\chi)$. For each bipartition of the MPS, the corresponding entanglement entropy is interpreted differently.
First, the bond connecting the spatial and temporal tensors captures the separability of variables $x$ and $t$. This is particularly insightful because the notion of separability is fundamental to PDEs. For linear PDEs, such as the heat equation, the separation of variables yields analytical solutions. Even for nonlinear PDEs, separability remains a powerful tool for finding analytical solutions in specific cases~\cite{Polyanin_Zhurov_2021}.
For separable solutions, the bonds connecting the spatial tensors determine the inter-scale correlations in space, and similarly, the bonds among the temporal tensors capture the interaction between slow and fast dynamics. 
However, for non-separable solutions, the bonds in the spatial and temporal parts carry correlations across the whole MPS chain. These cases are harder to characterize.

\begin{figure}[t!]
    \centering
    \includegraphics[width=0.9\linewidth]{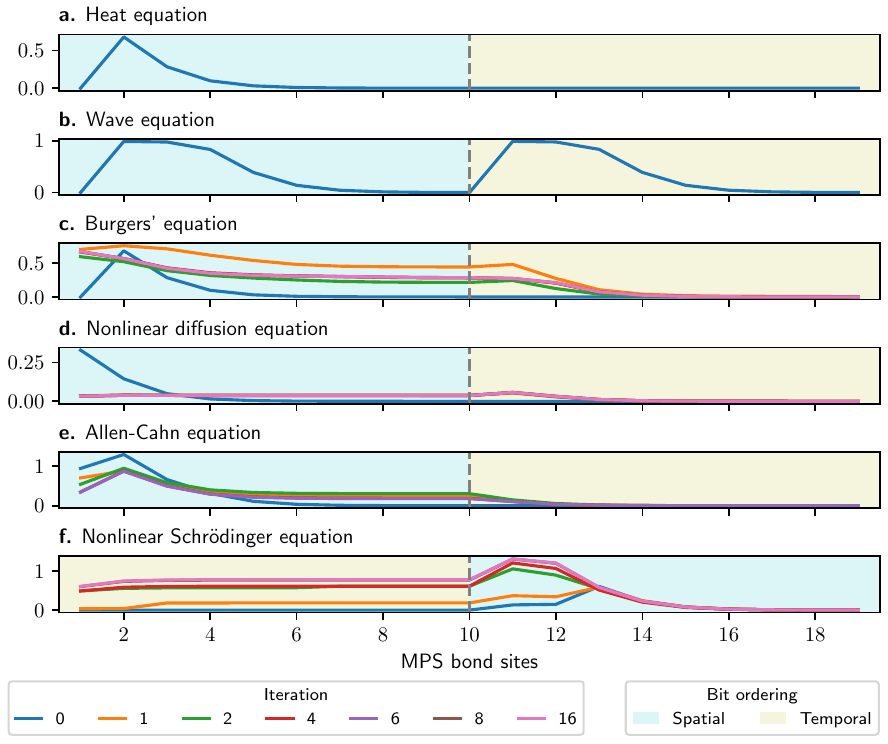}
    \caption{\textbf{Bipartite entanglement entropy of the MPS space-time solution}. We present the bipartite entanglement entropy as a measure of spatio-temporal correlations captured by the MPS ansatz. Each subplot corresponds to the specific cases summarized in Table~\ref{tab:equations}, and the details of each simulation are presented in Appendix~\ref{app:benchmarks}. For linear equations, we plot the entanglement of the final solution. For nonlinear equations, we plot the intermediate iterations of the Picard scheme. 
    For all PDEs, we use a 20-qubit MPS encoding to represent the solution on a $1024\times1024$ space-time grid. We exceptionally use the time-space ordering for the nonlinear Schr{\"o}dinger equation, motivated by an improvement in the convergence of the MPS solver that is described in Appendix~\ref{app:NLSE}.
    }
    \label{fig:entanglement}
\end{figure}

In Fig.~\ref{fig:entanglement}, we present the entanglement entropy of the MPS space-time solution for the PDEs summarized in Table~\ref{tab:equations}. For the heat equation in panel \textsf{a}, we observe that the exponential decay with time results in a separable temporal part, and the independence of the $x$ and $t$ variables is captured by zero entanglement at the space-time bond. In panel \textsf{b}, we observe that the spatial correlations exhibit a similar trend to the temporal part. This captures the equivalence between the $x$ and $t$ variables in the wave equation and its solution. We also see that the space-time parts are weakly entangled, as the solution remains separable along the characteristic coordinates of $x \pm ct$. For nonlinear equations, in panels \textsf{c}-\textsf{f}, we notice an immediate jump in entanglement between the spatial and temporal parts of the MPS. Despite this, the bond dimension remains small ($\chi \leq 32$), resulting in the number of MPS parameters being less than $1\%$ of the grid size. For both Burgers' and nonlinear diffusion equations, we observe that the temporal entanglement decays quickly. We understand this behavior as the manifestation of smooth evolution trajectories, shown in the Appendix~\ref{app:numerical_results}. For the 1D nonlinear Schr{\"o}dinger equation, we note that the temporal entanglement in this case does not decay but remains almost constant, which is a result of the \textit{time-space ordering} of the MPS. The change in index ordering is motivated by the observed improvement in the convergence of the iterative solver. We discuss this further in Appendix~\ref{app:NLSE}.

\paragraph{Convergence and runtime:}
The MPS representation is not only memory-efficient but also enables linear algebraic operations that scale logarithmically with grid size and polynomially with bond dimension~\cite{oseledets2011tensortrain}. We use DMRG-inspired solvers to solve the linear systems within the compressed representation~\cite{oseledets2012solution,dolgov2014alternating}.
For non-linear systems, we use Picard linearization to reduce the nonlinear problem to an iterative linear solve. Picard linearization is a fixed-point iteration scheme whose convergence is not guaranteed for a general nonlinear problem~\cite{Golub1992}. In our simulations, 16 iterations are sufficient to reach the errors shown in Table~\ref{tab:equations}. We can improve this convergence by using the Newton method to linearize the problem. In an accompanying work~\cite{ku_st_paper}, we present an MPS-based space-time solver using the Newton method along with benchmarks across several nonlinear regimes. Additionally, we use the multigrid method to improve the conditioning of the Jacobian at large grid sizes.

\begin{figure}[t!]
    \centering
    \includegraphics[width=0.9\linewidth]{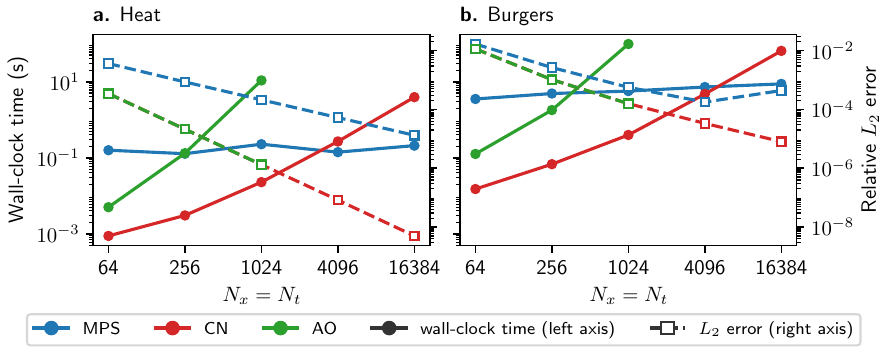}
    \caption{\textbf{Runtime and convergence under grid refinement at fixed $\chi_{max}=20$.} For the (1+1)D Heat and Burgers' equations, we present the scaling of runtime (solid lines) and convergence of the numerical error of approximation (dashed lines) against the refinement of the uniform grid. The space-time grid is refined by a factor of 4 in both dimensions, resulting in four additional MPS tensors ($L_x$ and $L_t$ increase by 2). We compare the MPS space-time solver against two classical numerical methods. \texttt{CN} is a time-stepping method using the Crank-Nicolson scheme, and \texttt{AO} is a classical space-time method that solves an all-at-once system like the MPS solver. For runtime comparison, we observe that the MPS solver stays flat while the classical methods scale poorly with grid refinement for both PDEs. On the other hand, the approximation error of classical solvers follows second-order convergence while the MPS solver follows linear convergence. This is expected as the MPS solver uses backward Euler approximation for the time derivatives and can be generalized to higher orders. We refer to the main text for an explanation of the deviation at $N_x=N_t=16,384$ in the case of Burgers' equation.}
    \label{fig:grid-refinement}
\end{figure}

In Fig.~\ref{fig:grid-refinement}, we compare the MPS space-time solver with a classical time-stepping solver (CN) and classical all-at-once method (AO), both based on the Crank-Nicolson discretization of time derivatives.
The classical methods are faster than the MPS solver on the coarsest grids, while the MPS solver is faster on the finest grids, and the runtime of the MPS solver grows slowly with the grid size. The classical all-at-once solver (AO) scales worst and is not run beyond $2^{10}$ points because of its memory use. On the other hand, the classical methods exhibit a quadratic convergence under grid refinement, while the MPS solver is only linear. This is due to the backward Euler scheme used in the MPS space-time solver. The MPS solver can be generalized to higher orders, and we consider this out of scope here. For the heat equation, the convergence follows the expected trend, but for the Burgers' equation, the MPS solver converges until $L_x=L_t=12$ and deviates slightly at $L_x=L_t=14$. Since the bond dimension is capped at 20 for all grids, the deviation is due to the amplification of MPS compression error during the Picard iterations. For other nonlinear PDEs, we observed that Picard iteration becomes unstable and the MPS solver at a fixed bond dimension shows no convergence under grid refinement. However, we can recover the convergence by allowing for higher bond dimensions at larger grid sizes. 

\paragraph{Complexity assumptions:}
The logarithmic scaling of the MPS solver with $N_x$ and $N_t$ holds at a fixed bond dimension $\chi$. One DMRG sweep grows linearly with the number of qubits $L_x+L_t$ and polynomially with $\chi$. We do not study how $\chi$ depends on the grid size, the parameters of the PDE, the time horizon, the ordering of the qubits, or the initial condition. These studies are out of scope for this work, and all scaling statements are conditional on a bounded $\chi$. For the simulations of Table~\ref{tab:equations}, $\chi_{\max}\le32$ on the $2^{10}\times2^{10}$ grid.

\subsection{MPS-DMD}

\begin{figure}[t!]
    \centering
    \includegraphics[width=0.95\linewidth]{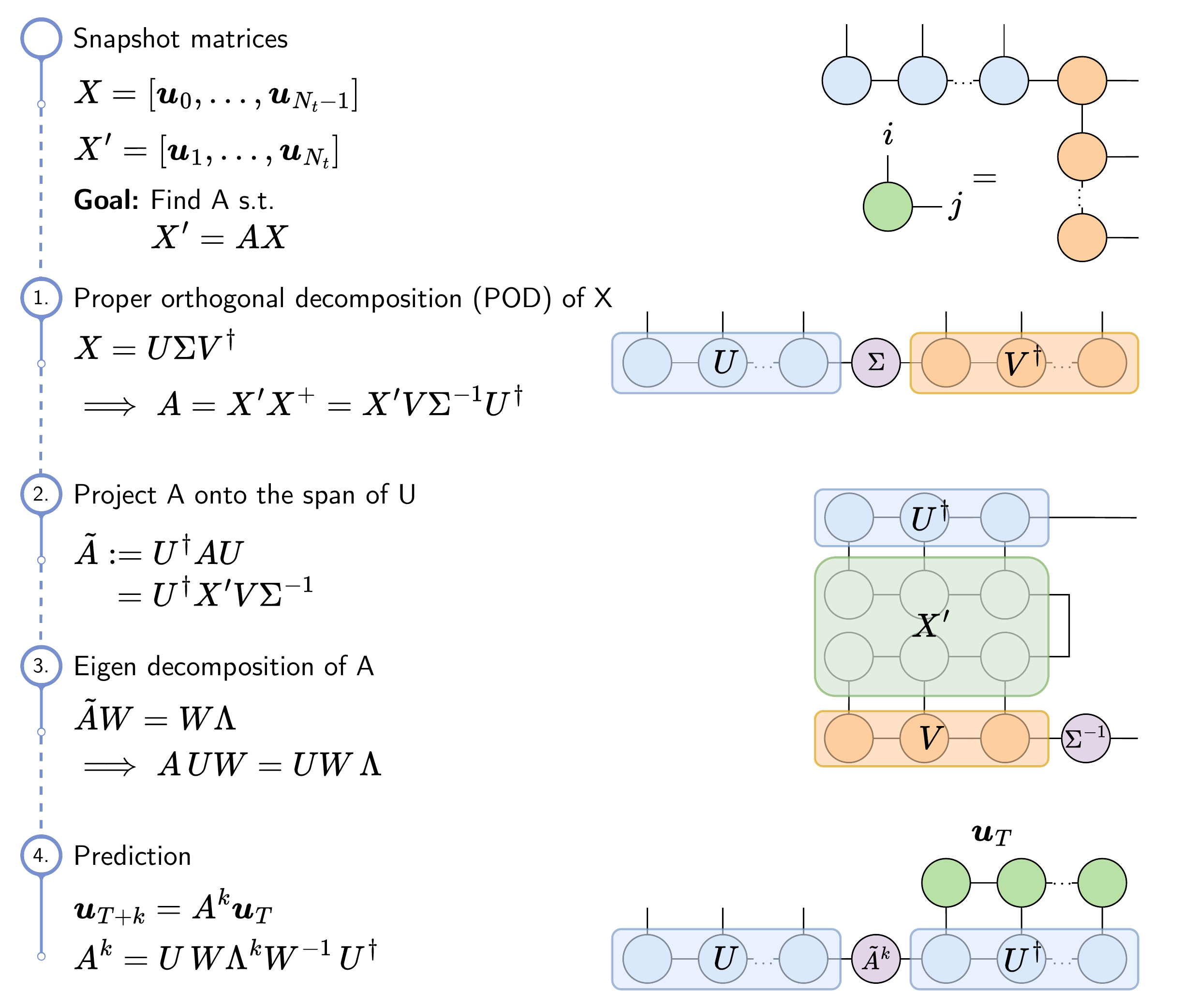}
    \caption{
    \textbf{MPS-DMD algorithm}. A graphical overview of the steps in the MPS-DMD algorithm for predicting future dynamics. At each step, we present the corresponding tensor network diagrams in Penrose notation. 
    We assume that the matrices $X$ and $X'$, containing snapshots $\{\boldsymbol{u}_j\}_{j=0}^{N_t}$, are provided in their MPS representations. First, we compute the proper orthogonal decomposition (POD) of $X$ by bringing the corresponding MPS into the mixed canonical form at the bond connecting the spatial and temporal parts. Now, the pseudoinverse of $X$, denoted by $X^+$, can be computed by inverting the singular values $\Sigma$. 
    We then project A$ A$onto the span of $U$ to obtain the reduced evolution operator $\tilde{A}$, which is used to efficiently compute the eigen decomposition of the full operator $A$. We further use this decomposition to predict the future dynamics.
    All matrix operations, except for the eigen-decomposition of $\tilde{A}$, are computed using only local contractions, resulting in logarithmic scaling with both the spatial and temporal resolution at a fixed bond dimension.
    }
    \label{fig:mps-dmd}
\end{figure}

Here, we develop the MPS-DMD algorithm. We refer to Section~\ref{sec:dmd} for an overview of the standard DMD algorithm~\cite{SCHMID_DMD_2010,kutz2016dynamic}. The time-series data is arranged into snapshot matrices, as shown in Fig.~\ref{fig:mps-dmd}. 
We assume that the snapshot matrices are provided in their corresponding space-time MPS. Later, we discuss efficient methods to convert arbitrary data into the required MPS form. Next, we see that the \textit{gauge freedom} of the MPS representation can be exploited to compute the proper orthogonal decomposition (POD) of $X:=U\Sigma V^{\dagger}$. We achieve this by transforming its MPS into the mixed canonical form~\cite{orus2014practical}, centered at the space-time bond. This is equivalent to the Schmidt decomposition from the Quantum Physics literature. 
As shown in step 1 of Fig.~\ref{fig:mps-dmd}, the isometries $U$ and $V^{\dagger}$ are given by the corresponding MPS tensors after the gauge transformation. 
And $\Sigma$ is the singular value matrix at the space-time bond of the MPS.

We then compute the (approximate) evolution operator $A=X'X^+\in \mathbb{R}^{N_x\times N_x}$. Here, $X^+$ is the pseudoinverse of $X$, which is obtained directly from its POD and leads to the decomposition $A=X'V\Sigma^{-1}U^{\dagger}$, with $\Sigma^{-1}$ the inverse of $\Sigma$. To efficiently diagonalize $A$, we compute the eigen decomposition of $\tilde{A}:=U^{\dagger}AU$, which is the projection of $A$ onto the span of $U$. $\tilde{A}$ is more convenient than $A$ since the size is $\chi \times \chi$ (See step 2 of Fig.~\ref{fig:mps-dmd}).
The resulting eigen decomposition is $\tilde{A}\,W = \Lambda\,W$, with eigenvalues $\Lambda$ as a diagonal matrix and eigenvectors $W$ in $\mathbb{C}^{r\times r}$. To compute the dynamic modes $\Phi$, we project $W$ back into the $x$-dimension to obtain $\Phi = UW$. However, we do not need explicit dynamic modes to predict dynamics.
Instead, we predict the state at future time step $k$ from the eigen decomposition of $\tilde{A}$ as:
\begin{equation}
\begin{aligned}
    \boldsymbol{u}_{T+k} :&=A^k \boldsymbol{u}_{T} 
    = U\tilde{A}^kU^{\dagger}\boldsymbol{u}_{T}\\
    &= U\, W\,\Lambda^k\,W^{-1}\, U^{\dagger} \boldsymbol{u}_{T} \ ,
\end{aligned}    
\end{equation}
where $\boldsymbol{u}_{T}$ is the spatial snapshot at time $T$. 
Since all matrix multiplications are achieved through local tensor contractions, the computational complexity of the MPS-DMD algorithm scales logarithmically with both $N_x$ and $N_t$. This is in contrast to Ref.~\cite{klus2018tensor}, which has runtime scaling linearly with the number of snapshots $N_t$. 

\begin{figure}[t!]
    \centering
    \includegraphics[width=0.9\linewidth]{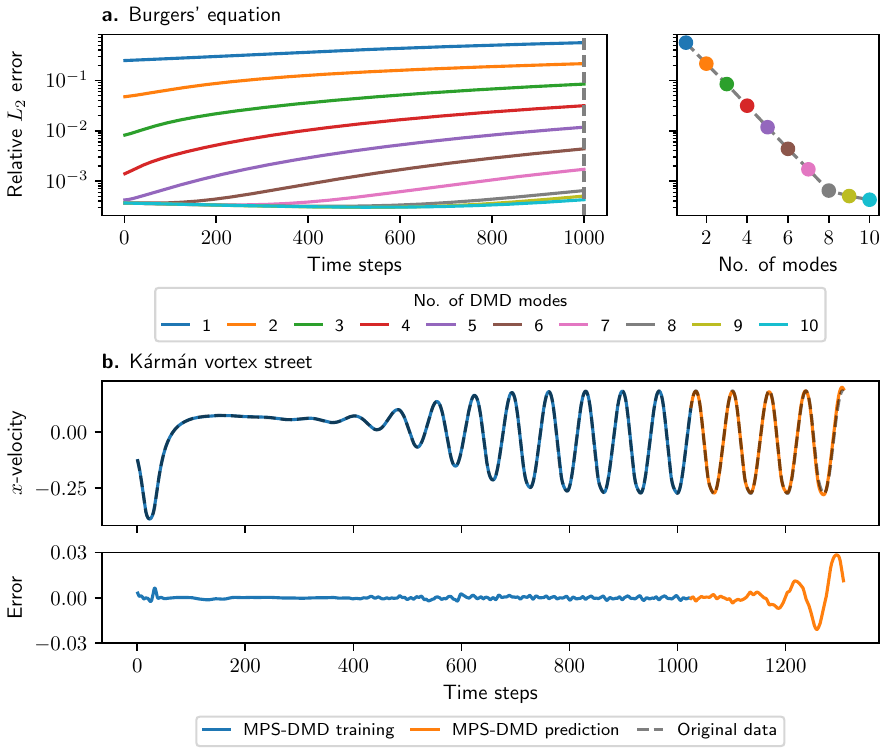}
    \caption{\textbf{MPS-DMD prediction of nonlinear dynamics.} 
    We look at two different applications of the MPS-DMD algorithm. For the Burgers' equation, the snapshot matrices are computed using the MPS space-time solver. For the Karman vortex street case, we compress dense snapshot matrices into the space-time MPS, i.e., the training phase of MPS-DMD.
    (\textsf{a.}) \textbf{1D Burgers' equation:}  We compute the snapshot matrices from the MPS space-time solution of the Burgers' equation, as described in Table~\ref{tab:equations}. We predict the solution for the next $1000$ time steps using the MPS-DMD algorithm. We plot the $L_2$ error of the predicted solution against the analytical solution for different truncations of the space-time bond dimension of MPS, which corresponds to the number of DMD modes.
    (\textsf{b.}) \textbf{2D K\'arm\'an vortex street phenomena:} We compress time-series data from Ref.~\cite{dataset_cylinder} into the MPS space-time representation. Specifically, we encode the time series of the x-velocity component using an \textit{interleaved} encoding (see Refs.~\cite{Erika_vlasov_eq,pisoni2025compression}) for the 2D space, but maintain the space-time ordering of the MPS. Using the proposed compression method, we build the space-time MPS with 25 qubits (15 in space and 10 in time) and bond dimension $\chi=200$. We then use MPS-DMD to predict the future dynamics and restrict the number of DMD modes to 100. We plot the x-velocity in the wake of the cylinder to capture the vortex shedding. The error plot is the difference between MPS-DMD predictions and reference data. The training error refers to the error due to MPS compression.
    }
    \label{fig:dmd_prediction}
\end{figure}

\paragraph{Compression methods:}\label{sec:compression} We discuss how to obtain the space-time MPS representation of the snapshot matrices.
First, using the MPS space-time solver, we can obtain the snapshot matrices directly from the MPS solution. Specifically, given an initial condition $\boldsymbol{u}_0$, we obtain the MPS space-time solution on the domain $\Omega\times(0, T]$ using the proposed solver. This MPS solution is used as the snapshot matrix $X'$ in the DMD method. Then, $X$ is computed by applying the downshift operator as an MPO on $X'$ and concatenating the initial condition $\boldsymbol{u}_0$. If the chosen time window of the MPS space-time solver captures all the relevant dynamics, then the DMD method can predict future solutions accurately.

We also consider the general case where the data is not available in the MPS format. In this case, if the data is known to have a low-rank MPS structure, then the benefits of MPS-DMD far outweigh the overhead of finding the MPS representation. In practice, we observe that the bond dimensions are small, consistent with the low temporal entanglement shown in Fig.~\ref{fig:entanglement}. 
Thus, we can obtain the MPS efficiently using methods such as optimized TT-SVD algorithms~\cite{rohrig2022}, TT-cross approximation~\cite{OSELEDETS201070TTcross}, or streaming TT decompositions~\cite{kressner2023streaming}. 

In particular, our method to compress an arbitrary data set is a streaming-like algorithm that is described next. The snapshot matrix $X'$ of size $N_x\times N_t$ is seen as:
\begin{equation}
    X' = \sum_{i=1}^{N_t} \boldsymbol{u}_i \otimes \boldsymbol{e}_i
\end{equation}
where $\boldsymbol{e}_i$ are the unit basis vectors for $i=1,\dots,N_t$.
We compress each $\boldsymbol{u}_i$ snapshot of the data into MPSs of $\log_2(N_x)$ tensors, using TT-SVD. Next, we compute the space-time MPS by sequentially adding the snapshots stacked along with the corresponding MPS encoding of $\boldsymbol{e}_i$, which has $\log_2(N_t)$ tensors and a bond dimension $\chi=1$. By truncating the bond dimension after each addition, we efficiently find the space-time MPS with $\log_2(N_x)+\log_2(N_t)$ tensors.
We remark here that our compression procedure scales linearly with the number of timesteps, while the standard TT-SVD compression~\cite{oseledets2011tensortrain} is cubic in the worst case.

\begin{figure}
    \centering
    \includegraphics[width=0.9\linewidth]{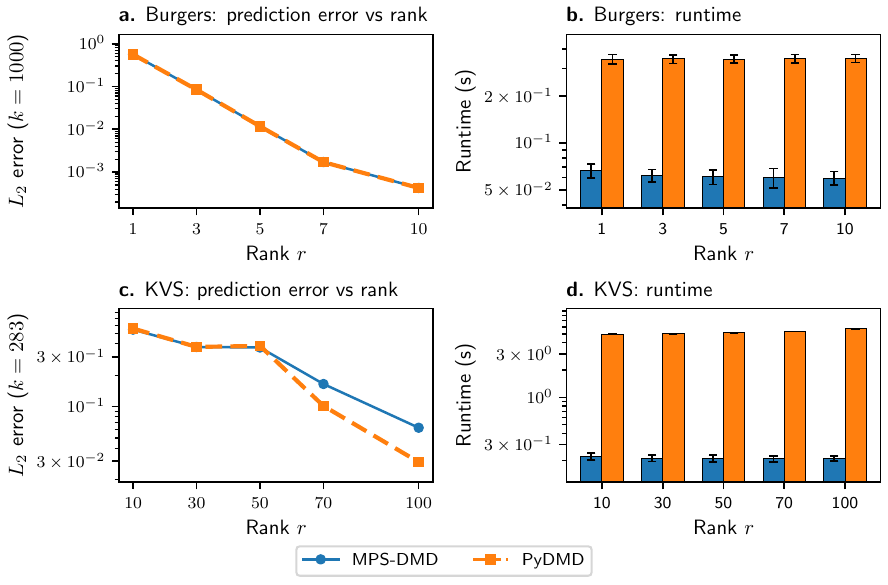}
    \caption{\textbf{Benchmark of MPS-DMD against standard PyDMD.}
    We compare the runtime and prediction error of the MPS-DMD algorithm and exact DMD of \texttt{PyDMD} for both Burgers' equation and the 
    Karman vortex street (KVS) phenomenon. We repeat the runtime measurement 10 times and plot the mean and standard deviation. For the case of Burgers' equation, the source of the time series is the MPS space-time solution, and we see no effect of truncation error as the data is low-rank. For the KVS data, the space-time MPS is computed from classical simulation data, and we note that the compression step introduces an error that affects the DMD predictions at higher ranks. However, this error is purely due to the compression method and not systematic of the MPS-DMD method. 
    }
    \label{fig:mps-dmd-benchmark}
\end{figure}

\paragraph{Numerical results:}
In Fig.~\ref{fig:dmd_prediction}, we compute the MPS-DMD predictions for two different nonlinear systems. We present results using both the compression methods discussed above. In panel \textsf{a.},  we start from the MPS space-time solution on a $1024 \times 1024$ grid for the (1+1)D Burgers' equation. Using the 20-bit MPS with $\chi=10$ obtained in the previous section, we perform MPS-DMD predictions for different truncations of the space-time bond dimension, corresponding to varying numbers of DMD modes. We observe that the prediction error remains small for up to $k = 1000$ time steps into the future, despite the MPS representation having a bond dimension of only 10, resulting in a compression of over $99\%$. In Appendix~\ref{app:dmd_horizon}, we observe that MPS-DMD predictions match those of exact DMD up to $k=10^5$ future time steps without any relearning.
In panel \textsf{b.}, we compress the simulation data from Ref.~\cite{dataset_cylinder} into a space-time MPS. Specifically, we sequentially convert each of the $1024$ snapshots of the x-velocity component (on a grid of $256\times128$ points) into 15-bit MPSs. We combine them to compress the (1+2)D velocity snapshot matrix into a 25-bit MPS with a bond dimension $\chi=200$, reducing memory usage by $97.42\%$. We observe that the MPS-DMD algorithm accurately predicts the Karman vortex street phenomenon using only 100 DMD modes.
In Fig.~\ref{fig:mps-dmd-benchmark}, we compare MPS-DMD against exact DMD of the \texttt{PyDMD} library for the above two cases. We measure the runtime of both algorithms, repeated 10 times for each rank, starting from the same time-series data. For both cases, the measurements do not include the time taken to generate the time series data in the appropriate form. We see that MPS-DMD is roughly an order of magnitude faster with almost identical accuracy in prediction, for both Burgers' and Karman vortex street dynamics.

\section{Conclusions}\label{sec:conclusions}
We investigated the effectiveness of the MPS representation in the combined space-time domain for both simulation and data-driven prediction.
For both linear and nonlinear (1+1)D PDEs, we observed that the MPS space-time solution captures the key spatio-temporal correlations with small bond dimensions, yielding compression ratios exceeding $99\%$ on a $2^{10}\times2^{10}$ grid.  We also extend the solver to (2+1)D PDEs and solve the 2D Heat, Burgers', and Allen-Cahn equations.
Our tensor network solver can also be extended to develop space-time methods on quantum computers~\cite{PhysRevResearch.6.033257}.
We introduced an MPS version of the well-known DMD method that successfully predicts the long-term dynamics of nonlinear systems. Under the assumption of low spatio-temporal entanglement,i.e., bounded bond dimension, this has an exponentially lower computational cost than standard DMD. %
A direct future direction would be to study the MPS space-time encoding for higher-dimensional and vector-valued PDEs. 
Interestingly, the MPS ansatz natively imposes a hierarchical decomposition of the timescales, inducing a separation between the fast and slow dynamics.
This separation may be explored further to improve DMD predictions~\cite{kutz2016multiresolution}. 
Our findings could also lead to interesting real-world applications in aerodynamics and weather forecasting, where nonlinear dynamics are dominant.

\section*{Acknowledgements}
RP, SP, ET, and LA thank Jonas Fuska and Ingo Roth for fruitful discussions. We also thank the anonymous reviewers for providing valuable feedback.

\paragraph{Funding information}
NR, YA, and MKR gratefully acknowledge support from grant number KUEX-8434000491, a collaborative effort between Khalifa University, the Technology Innovation Institute, and the Emirates Nuclear Energy Company. The Emirates Nuclear Technology Center is a collaboration among Khalifa University of Science and Technology, Emirates Nuclear Energy Company, and the Federal Authority for Nuclear Regulation.

\begin{appendix}
\numberwithin{equation}{section}
\section{Space-time MPS formulation}\label{app:space-time}
In this appendix, we present the construction of the \textit{all-at-once} space-time linear system starting from the time-stepping equation. We also describe the corresponding MPS implementation.
First, we recall the standard (1+1)-dimensional time-dependent PDE of $u$:
\begin{equation}
    \partial_t^p u + F(u,\partial_x u,\dots,\partial_x^{q} u) = g(x,t)
\end{equation}
For clarity, we stick to the first-order derivative in time, $p=1$. We further set $g(x,t)=0$. 
We start with the nonlinear case, i.e., where $F$ is dependent on the solution $u$. Since the linear case is a simplified version, it follows straightforwardly. 

After the discretization of the domain, we represent the discretized operator version of $F$ with $\mathbb{F}(u)$ acting on the solution $u$. Then, for the implicit scheme, the time-stepping equation at time step $j$ is given by:
\begin{equation}\label{eq:nonlin_pde_eq_imp}
    \frac{u_{i,j+1}-u_{i,j}}{\tau} + \mathbb{F}(u_{i,j+1}) u_{i,j+1} = 0 .
\end{equation}
We define the implicit operator $\mathbb{O} = \mathbb{I} + \mathbb{F}(u) \tau$, such that Eq.~\eqref{eq:nonlin_pde_eq_imp} is rewritten as $\mathbb{O}(u) \boldsymbol{u}_{j+1} = \boldsymbol{u}_{j}$, where $\boldsymbol{u}_j$ is a column vector containing the solution at all spatial grid points and fixed time step $j$. We linearize the equation using a Picard iterative update, where at iteration $p$ the linear operator is given by substituting the solution at iteration $p-1$, i. e. $\mathbb{O}(\boldsymbol{u}^{p-1}_{j+1})\boldsymbol{u}^p_{j+1} = \boldsymbol{u}^p_j$. For the sake of brevity, we now refer to the linearized operator simply as $\mathbb{O}_x$.

We now look at the linear system for the entire space-time solution vector $\boldsymbol{u}$, given by the tensor product:
\begin{equation}\label{eq:implicit_tensorization}
    (\mathbb{O}_x \otimes \mathbb{I}_t)\boldsymbol{u} = \boldsymbol{u}^{\mathrm{shifted}} \,,
\end{equation}
where $\mathbb{I}$ is the identity operator and the subscripts $x$ and $t$ are used to emphasize the spatial and temporal operators.
$\boldsymbol{u}^{\mathrm{shifted}}$ denotes the space-time solution shifted backward by one time step $\tau$. 
More explicitly, $\boldsymbol{u}$ and $\boldsymbol{u}^{\mathrm{shifted}}$ are given by the element-wise stacking of vectors $\boldsymbol{u}_j$:
\begin{equation}
\begin{aligned}
    \boldsymbol{u} &= (u_{1,1}, u_{1,2}, ..., u_{1,N_t}, u_{2,1}, ..., u_{N_x,N_t})^T  \\
    \boldsymbol{u}^{\mathrm{shifted}} &= (u_{1,0}, u_{1,1}, ..., u_{1,N_t-1}, u_{2,0}, ..., u_{N_x,N_t-1})^T
\end{aligned}
\end{equation}
Additionally, $\boldsymbol{u}^{\mathrm{shifted}}= (\mathbb{I}_x \otimes \mathbb{S}_t) \boldsymbol{u} + \boldsymbol{u}_{0} \otimes \boldsymbol{e}_1$ where $\boldsymbol{e}_{1}\in \mathbb{R}^{N_t}$ is the one-hot vector with value one in the first position, and the shift operator $\mathbb{S}_t$ is:
\begin{equation}
    \mathbb{S}_t = 
    \begin{pmatrix}
        0 & 0 & 0 & 0 \\
        1 & 0 & 0 & 0 \\
        0 & 1 & 0 & 0 \\
        0 & 0 & 1 & 0 \\
        & & & \ddots & \ddots
    \end{pmatrix}_{N_t \times N_t} .
\end{equation}
Finally, we build the \textit{all-at-once} space-time linear system as:
\begin{equation}\label{eq:space_time_linsys}
    (\mathbb{O}_x \otimes \mathbb{I}_t - \mathbb{I}_x \otimes  \mathbb{S}_t) \boldsymbol{u} = \boldsymbol{u}_{0} \otimes \boldsymbol{e}_1 \ .
\end{equation}
We remark that this is equivalent to the all-at-once linear system presented in the main text. We now translate all the vectors and operators in Eq.\eqref{eq:space_time_linsys} into the MPS-MPO formalism. 

The operator $\mathbb{O}_x$ arises from spatial derivatives and functions of the solution in the case of nonlinear equations. The spatial derivatives require Toeplitz matrices, for which analytical MPO constructions are known~\cite{Kazeev2013}. All other matrices, such as identity or shift matrices, can also be viewed as simple Toeplitz matrices with a single diagonal. The nonlinear functions of $\boldsymbol{u}$ can either be constructed from the powers of the MPS compressed with TT-SVD or using TT-cross interpolation directly from the MPS representing the solution~\cite{Savostyanov2011FastAI,ritter2024quantics}. The tensorization in Eq.\eqref{eq:space_time_linsys} results in the concatenation of the constituent MPOs (and MPSs) for the spatial and temporal parts. This completes the assembly of the all-at-once linear system in the MPS format, which is then solved using DMRG-inspired algorithms. DMRG is a variational method that optimizes the MPS tensors locally, obtaining the MPS solution with a runtime scaling polynomially with the bond dimension $\chi$, ranging from $\mathcal{O}(\chi^3)$ to $\mathcal{O}(\chi^6)$ depending on the variant~\cite{oseledets2012solution,dolgov2014alternating}.

\paragraph{Boundary conditions and nonlinear terms:}
For implementing Dirichlet boundary conditions, we place the $2^{L_x}$ grid nodes such that the zero ghost values of the Toeplitz difference MPOs fall on the domain boundary for homogeneous conditions. Inhomogeneous values enter the right-hand side as a boundary-condition MPS.
Nonlinear coefficients are diagonal MPOs built from the current iterate with MPS-MPO products and by TT-cross interpolation. 

\paragraph{Two spatial dimensions:}
For (2+1)D equations, the spatial MPOs act on $L_x+L_y$ qubits, and the space-time MPS has $L_x+L_y+L_t$ qubits. We use \textit{interleaved}, $[x_1,y_1,\dots,x_L,y_L,t_1,\dots,t_{L_t}]$, or \textit{stacked}, $[x_1,\dots,x_L,y_1,\dots,y_L,t_1,\dots,t_{L_t}]$, ordering of the spatial qubits. The operators are sums of Kronecker products of the one-dimensional ones, e.g., $\Delta=\partial_{xx}\otimes\mathbb{I}_y+\mathbb{I}_x\otimes\partial_{yy}$.

\paragraph{Implementation and hardware:}
All experiments were run on 32 threads of an Intel Core i9-13900K CPU with the \texttt{tensnet}, our in-house tensor network library built on top of PyTorch.

\section{Numerical results}\label{app:numerical_results}
Here, we present the detailed numerical simulations using our MPS space-time solver, following the order and the settings of Table~\ref{tab:equations}. In Sec.~\ref{app:benchmarks}, we describe the (1+1)D benchmarks, grouped by type of equation: linear, Burgers', Allen-Cahn, and nonlinear diffusion and nonlinear Schr{\" o} dinger equations. For each equation, we start from a case where the solution is known, or the problem is dominated by diffusion, and, where Table~\ref{tab:equations} has one, we add a second, harder regime that is not dominated by diffusion. In Sec.~\ref{app:benchmarks_2d}, we present the (2+1)D benchmarks.

Unless stated otherwise, the space-time domain is discretized by $2^{10}\times2^{10}$ points, encoded in a 20-qubit MPS, and all nonlinear equations are run for 16 Picard/Newton iterations with 4 DMRG sweeps per iteration. All the figures of Sec.~\ref{app:benchmarks} share the same layout. Panel \textsf{a.} shows the MPS space-time solution ($u_\mathrm{mps}$), with the spatial dimension ($x$) along the vertical and the time ($t$) along the horizontal axis. Panel \textsf{b.} shows the deviation ($u_\mathrm{mps}-u^*$) from the reference solution ($u^*$), which is analytical where available and otherwise obtained by classical time stepping. Panel \textsf{c.} shows the bond entanglement entropy of the MPS solution at each bipartition of the chain, with the spatial and temporal qubits shaded and the intermediate Picard/Newton iterates drawn for nonlinear equations.

\subsection{1D Benchmarks}\label{app:benchmarks}

\subsubsection*{Linear equations}
\paragraph{Heat equation:}
The heat equation, or diffusion equation, $u_t-\alpha u_{xx} = 0$, is a parabolic second-order PDE that governs many interesting phenomena. It is well-posed as an initial-boundary value problem. Here, we consider the domain $\Omega=(0,2\pi)\times(0, T]$ with $\alpha=1$, the initial condition $u_0(x)=\sin(x)$ and Dirichlet boundary conditions. The analytical solution is $u^*(x,t)= u_0(x)e^{-\alpha t}$. We encode the space-time solution for an evolution time of $T=1024 \tau$, where $\tau = 10^{-3}$. We use backward Euler in time. A bond dimension of $\chi=2$ suffices, as the solution is a single product of a spatial and a temporal function.

\paragraph{Wave equation:}
The 1D wave equation, $u_{tt}-c^2u_{xx}=0$, is a hyperbolic PDE that requires initial conditions on both $u$ and $u_t$ to be well-posed. We set $c=1$ and initialize the problem with $u(x,0)=\sin(\pi x)$ and $u_t(x,0)=0$ on the domain $\Omega = (0,10)\times(0, T]$ with homogeneous Dirichlet boundary conditions, where $T=1024\tau$ and $\tau = 5\times10^{-3}$. The analytical solution is the superposition of a left- and a right-moving wave,
\begin{equation}
    u^*(x,t)=\tfrac{1}{2}\left[\sin(\pi (x + ct))+\sin(\pi (x - ct))\right]=\sin(\pi x)\cos(\pi c t),
\end{equation}
which is of bond dimension $\chi=2$. The second-order time derivative uses the central difference $\partial_t^2\to(\mathbb{S}_t^{-1}-2\mathbb{I}+\mathbb{S}_t)/\tau^2$, with the spatial Laplacian averaged over three time levels with weights $(\theta,1-2\theta,\theta)$, $\theta=1/4$, which is the second-order Newmark scheme.

\begin{figure}[t]
\centering
    \begin{subfigure}{0.48\textwidth}
        \includegraphics[width=\linewidth]{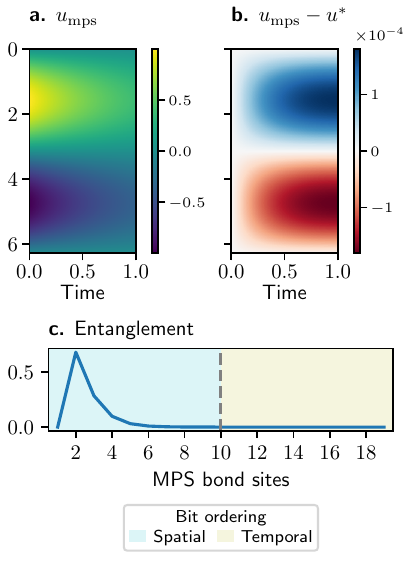}
        \caption{Heat equation $u_t-\alpha u_{xx} = 0$}
        \label{fig:heat_eqn}
    \end{subfigure}
    \hfill
    \begin{subfigure}{0.48\textwidth}
        \includegraphics[width=\linewidth]{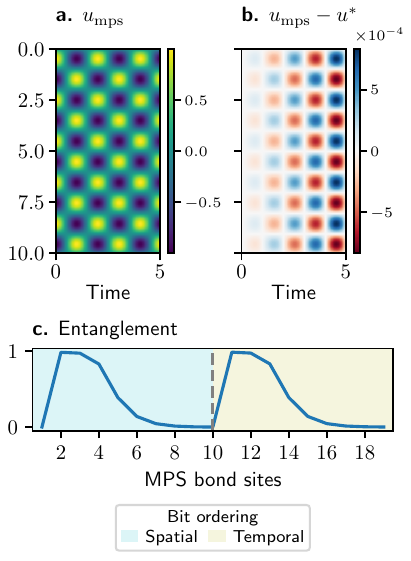}
        \caption{Wave equation $u_{tt}-c^2u_{xx}=0$}
        \label{fig:wave_eqn}
    \end{subfigure}
    \caption{\textbf{Linear PDEs.} MPS space-time solutions of the heat and wave equations, compared with their analytical solutions. For both, the bond dimension is $\chi=2$, and the entanglement at the bond between the spatial and temporal qubits is zero, reflecting the separability of the solutions in $x$ and $t$.}
    \label{fig:linear_benchmarks}
\end{figure}

\subsubsection*{Burgers' equation}
\paragraph{Smooth solution:}
In nonlinear problems, we first solve the 1D Burgers' equation, a fundamental equation that models the convection-diffusion process and can be viewed as a 1D version of the Navier-Stokes equations for a scalar variable. We use the MPS space-time solver for the linearized Burgers' equation, $u_t+uu_x-\nu u_{xx}=0$, and iterate until convergence; $\mathbb{F}(u)$ contains the diagonal MPO $\mathrm{diag}(u)\,\partial_x$, built from the previous Picard iterate. We benchmark the solver for a specific instance where the analytical solution is known~\cite{Wood_2006}, given by $u^*(x,t)=\frac{2\nu\pi e ^{-\nu\pi^2 t}\sin(\pi x)}{1.01+e ^{-\nu\pi^2 t}\cos(\pi x)}$ with $\nu=10^{-3}$. Setting $t=0$ yields the initial condition, and the homogeneous Dirichlet boundary conditions are $u=0$ at $x=0$ and $x=1$. We solve on $\Omega=[0,1]\times(0,T]$ with $T=10$. The solution is smooth and diffusive, and a bond dimension of $\chi=14$ gives an $L_2$ error of $5.7\times10^{-4}$.

\paragraph{Shockwave formation:}
We then solve a more challenging regime of the 1D Burgers' equation, which is not dominated by diffusion and leads to the formation of a shockwave. We set the initial condition $u_0(x)= \sin(x)$ on $[0,2\pi]$, with homogeneous Dirichlet boundary conditions and a viscosity of $\nu=5\times10^{-3}$, for an evolution time of $T=8.192$. This is a regime for which the space-time method is better suited than time stepping, as the shock requires small time steps. The analytical solution is a series in Bessel functions that cannot be summed in double precision at this viscosity, so we compare against a pseudo-spectral solution. The MPS solution captures the shock with $\chi=20$, and the error is concentrated at the shock front.

\begin{figure}[t]
\centering
    \begin{subfigure}{0.48\textwidth}
        \includegraphics[width=\linewidth]{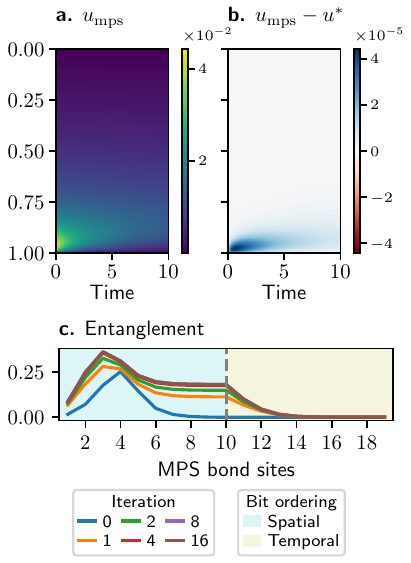}
        \caption{Smooth solution ($\nu=10^{-3}$)}
        \label{fig:burgers_smooth}
    \end{subfigure}
    \hfill
    \begin{subfigure}{0.48\textwidth}
        \includegraphics[width=\linewidth]{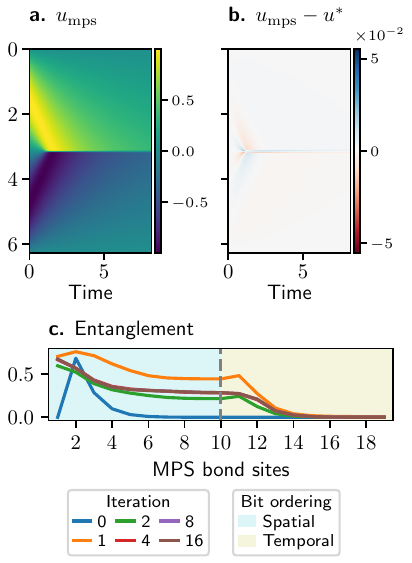}
        \caption{Shockwave formation ($\nu=5\times10^{-3}$)}
        \label{fig:burgers_eqn}
    \end{subfigure}
    \caption{\textbf{Burgers' equation.} MPS space-time solutions of $u_t+uu_x-\nu u_{xx}=0$ for the smooth case, compared with the analytical solution, and for the shockwave, compared with a pseudo-spectral solver. The entanglement plots show the Picard iterates. The shockwave case requires a higher bond dimension ($\chi=20$) and a larger entanglement than the smooth case ($\chi=14$).}
    \label{fig:burgers_benchmarks}
\end{figure}

\subsubsection*{Allen-Cahn equation}
\paragraph{Gaussian initial condition:}
The Allen-Cahn equation, $u_t=\epsilon^2u_{xx}+u-u^3$, is a reaction-diffusion equation that describes phase separation, with stable phases $u=\pm1$ separated by interfaces of width $\sim\epsilon$. There is no closed-form solution for our initial conditions, so we compare against an explicit finite-difference solver. We use backward Euler and Newton's method: the Jacobian contains the diagonal MPO $\tau\,\mathrm{diag}(3u^2-1)$ in place of the Picard operator. We first consider the domain $x\in[-1,1]$ with periodic boundary conditions, $\epsilon=0.05$, $T=0.1$ and a Gaussian initial condition $u_0(x)=e^{-x^2/0.18}$. In this case, the solution is almost purely diffusive, and no interfaces form during the evolution. Newton converges with $\chi=17$ to an $L_2$ error of $2.7\times10^{-5}$.

\paragraph{Spinodal initial condition:}
In the second case, the initial condition is a rough random field, $u_0(x)=\sum_{k=1}^{10}(a_k\sin\pi kx+b_k\cos\pi kx)/k$ with $a_k,b_k\sim\mathcal{N}(0,0.4)$, evolved for $T=4$ with $\epsilon=0.15$. The field separates into domains of $u=\pm1$, whose walls then move and annihilate, so that the number of domains decreases during the evolution. Here, the solution is not dominated by diffusion, and Newton's method is much more sensitive to the bond dimension: with $\chi=20$ the iteration stalls at a residual of order $5\times10^{-2}$. We therefore cap the bond dimension at $\chi=35$ and loosen the Newton tolerance to $10^{-3}$. The converged solution has $\chi_\mathrm{max}=32$ and an $L_2$ error of $3.6\times10^{-4}$.

\begin{figure}[t]
\centering
    \begin{subfigure}{0.48\textwidth}
        \includegraphics[width=\linewidth]{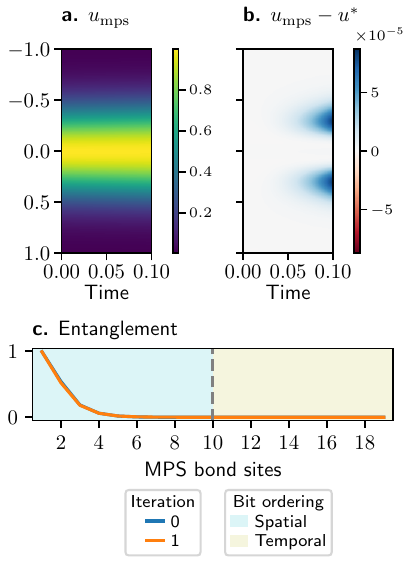}
        \caption{Gaussian initial condition ($\epsilon=0.05$)}
        \label{fig:allen_cahn_smooth}
    \end{subfigure}
    \hfill
    \begin{subfigure}{0.48\textwidth}
        \includegraphics[width=\linewidth]{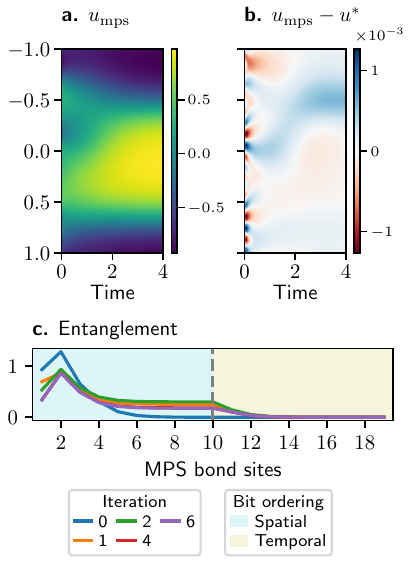}
        \caption{Spinodal initial condition ($\epsilon=0.15$)}
        \label{fig:allen_cahn_eqn}
    \end{subfigure}
    \caption{\textbf{Allen-Cahn equation.} MPS space-time solutions of $u_t=\epsilon^2u_{xx}+u-u^3$ with periodic boundary conditions, compared with an explicit finite-difference solver. The entanglement plots show the Newton iterates. In the spinodal case, the formation and the coarsening of domains require a roughly twofold larger bond dimension.}
    \label{fig:allen_cahn_benchmarks}
\end{figure}

\subsubsection*{Nonlinear diffusion and nonlinear Schr\"odinger equations}
\paragraph{Nonlinear diffusion:} We solve the nonlinear diffusion equation, a nonlinear variant of the heat equation that models phenomena in which the material properties depend on the system's temperature,
\begin{equation}
\rho c_p \frac{\partial u}{\partial t} = \frac{\partial}{\partial x} \left( \kappa(u) \frac{\partial u}{\partial x} \right)\ ,
\end{equation}
where $\kappa(u) = \kappa_0\exp(\lambda u)$ and $\rho c_p=1$. We choose the initial condition $u_0(x) = \frac{5-x}{2} + (x-1)(x-3)$ on $x \in [1,3]$ with the Dirichlet boundary conditions $u(1,t)=2$ and $u(3,t)=1$, for which analytical solutions are not available~\cite{Polyanin_Zaitsev_2016}. We take $\kappa_0 = 0.1$, $\lambda=0.5$ and $T=15$. The conductivity $\kappa(u)$ is obtained by TT-cross interpolation from the previous Picard iterate, and the problem is linearized by Picard iteration. We benchmark against a classical finite-difference solver that performs implicit time stepping~\cite{filipov2018implicit}. 

\paragraph{Non-linear Schr\"odinger (NLS) equation:}
We also consider the NLS equation, which arises in quantum physics and has applications ranging from condensed matter systems to wave mechanics. Specifically, we consider the 1D `focusing' case, $i \psi_t = - \psi_{xx} - 2 |\psi|^2 \psi$, where $\psi (x,t)$ is a complex-valued function for all $(x,t) \in [-3\pi,3\pi] \times (0,T=3]$ with periodic boundary conditions. This equation admits an analytical solution, known as the bright soliton, that we use to benchmark our space-time MPS solver (see Appendix~\ref{app:NLSE} for the initial condition and the time-space ordering of the qubits used here). We use Crank-Nicolson in time because backward Euler damps the soliton, and Picard iteration with the density $|\psi|^2$ as a diagonal MPO. We plot the density $|\psi|^2$ as the solution is complex-valued.

\begin{figure}[t]
\centering
    \begin{subfigure}{0.48\textwidth}
        \includegraphics[width=\linewidth]{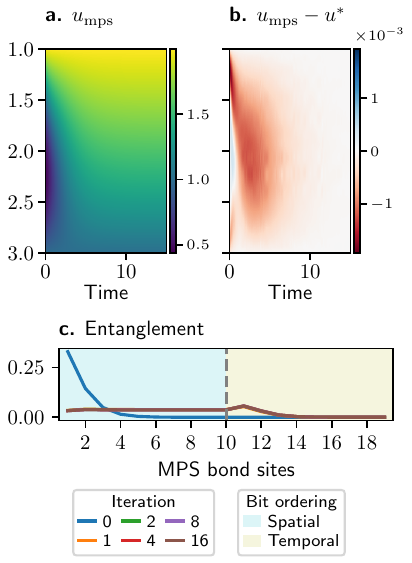}
        \caption{Nonlinear diffusion equation}
        \label{fig:nonlinear_diffusion}
    \end{subfigure}
    \hfill
    \begin{subfigure}{0.48\textwidth}
        \includegraphics[width=\linewidth]{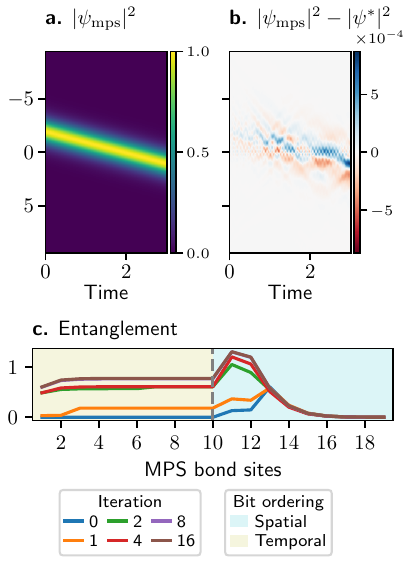}
        \caption{Nonlinear Schr\"odinger equation}
        \label{fig:nls_eqn}
    \end{subfigure}
    \caption{\textbf{Nonlinear diffusion and nonlinear Schr\"odinger equations.} (a)~The nonlinear diffusion equation is compared with a classical implicit finite-difference solver. (b)~The density $|\psi|^2$ of the NLS equation is compared with the analytical bright soliton. We use the time-space ordering for the MPS, as seen in the entanglement plot.}
    \label{fig:nl_benchmarks}
\end{figure}

\subsection{2D Benchmarks}\label{app:benchmarks_2d}
\begin{figure}[t]
    \centering
    \includegraphics[width=0.9\linewidth]{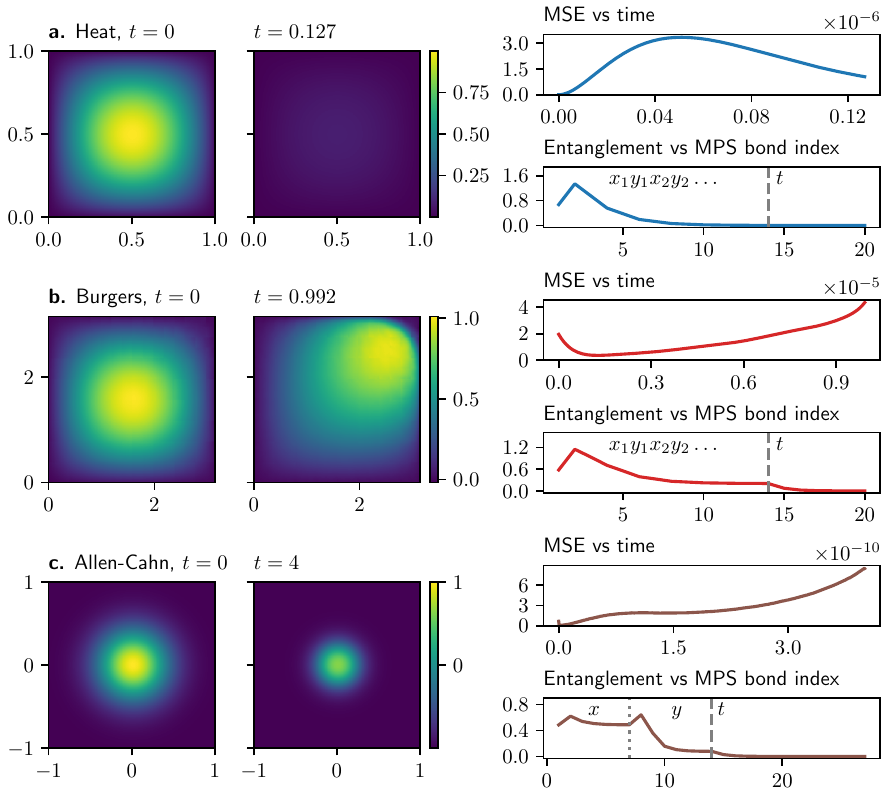}
    \caption{\textbf{Space-time MPS solutions of (2+1)-dimensional PDEs on a $2^7\times 2^7$ spatial grid}: (a) the heat equation, (b) the viscous Burgers' equation and (c) the Allen--Cahn equation. In each row, the two snapshots on the left show the MPS solution at $t=0$ and at the last time step. On the right, the top panel shows the MSE against a classical reference (analytical for the heat equation), and the bottom panel shows the bond entanglement entropy of the MPS. The dashed line marks the boundary to the temporal qubits and the dotted line the $x|y$ cut.}
    \label{fig:benchmark_2d}
\end{figure}

We now consider (2+1)-dimensional PDEs on a spatial grid of $2^7\times2^7$ points, and show the heat, Burgers', and Allen-Cahn equations in Fig.~\ref{fig:benchmark_2d}. The two snapshots on the left show the MPS solution at $t=0$ and at the last time step, on a shared color scale. On the right, the top panel shows the mean squared error (MSE) of the MPS solution against a classical reference, averaged over the spatial grid at each time, and the bottom panel shows the bond entanglement entropy of the final MPS, where the dashed line separates the spatial from the temporal qubits. The bond entropy is plotted against the cut position along the MPS chain. The labels give the order of the spatial qubits, and the dotted line marks the cut between the $x$ and $y$ qubits. In Table~\ref{tab:2d_summary}, we present the grid size, maximum bond dimension and the $L_2$-error for the MPS solution of each equation.

\paragraph{Heat equation:}
We solve $u_t=\Delta u$ on $[0,1]^2$ with Dirichlet boundary conditions and the initial condition $u_0=\sin\pi x\,\sin\pi y$, for $2^7$ time steps of $\tau=10^{-3}$. The spatial qubits are interleaved ($x_1y_1x_2y_2\ldots$). The analytical solution is $u^*=u_0\,e^{-2\pi^2t}$, so the solution decays by a factor of about 12 and its final snapshot looks nearly uniform on the shared color scale. The bond dimension is $\chi=4$, the $L_2$ error is $6.5\times10^{-3}$ and the residual is $2\times10^{-13}$. 

\paragraph{Burgers' equation:}
We solve the scalar viscous Burgers' equation, $u_t+u(u_x+u_y)=\nu\Delta u$, on $[0,\pi]^2$ with $\nu=10^{-2}$, Dirichlet boundary conditions and $u_0=\sin x\,\sin y$, up to $T=1$, using the interleaved ordering. There is no closed-form solution, so we compare against an explicit fourth-order Runge--Kutta finite-difference solver with substeps chosen to satisfy the CFL condition. We use the smallest bond dimension that gives an $L_2$ error below $10^{-2}$, which is $\chi=15$ (error $7.7\times10^{-3}$). 

\paragraph{Allen-Cahn equation:}
We solve $u_t=\epsilon^2\Delta u+u-u^3$ with $\epsilon=0.10$ on the periodic domain $[-1,1]^2$, for $T=4$ with $2^{14}$ time steps. The initial condition is a disc of the $+1$ phase in the $-1$ phase, $u_0=2\exp[-r^2/(2\sigma^2)]-1$ with $\sigma=0.3$. The interface sharpens, and the disc then shrinks by curvature-driven motion. Here, we use the stacked ordering of the spatial qubits ($x_1\ldots x_Ly_1\ldots y_L$). We compare against explicit finite differences. The bond dimension is capped at $\chi=40$, and we run 16 Newton iterations with a tolerance of $5\times10^{-4}$, which is loosened as in the spinodal case in 1D. 

\begin{table}[t!]
\centering
\small
\begin{tabularx}{0.9\textwidth}{XCCC}
\toprule
PDE & Grid $N_x\times N_y\times N_t$ & $\chi_{\max}$ & $L_2$-error \\
\midrule
Heat & $2^7\times2^7\times2^7$ & 4 & $6.5\times10^{-3}$  \\
Burgers' ($\nu=10^{-2}$) & $2^7\times2^7\times2^7$ & 15 & $7.7\times10^{-3}$  \\
Allen-Cahn ($\epsilon=0.10$) & $2^7\times2^7\times2^{14}$ & 40 & $1.9\times10^{-5}$  \\
\bottomrule
\end{tabularx}
\caption{\textbf{Summary of the (2+1)D simulations of Fig.~\ref{fig:benchmark_2d}.}}
\label{tab:2d_summary}
\end{table}

\section{MPS time-space encoding}\label{app:NLSE}

In this appendix, we discuss the time-space ordering of the MPS. In the MPS representation, the time-space encoding is expressed by:
\begin{equation}\label{eq:ts_mps}
    \boldsymbol{\psi} = A_1^{(t_1)}A_2^{(t_2)}\dots A_m^{(t_m)} A_{m+1}^{(x_1)}A_{m+2}^{(x_2)}\dots A_{m+n}^{(x_n)} \ .
\end{equation}
We consider this encoding exclusively for the (1+1)-dimensional NLSE, described by:
\begin{equation}\label{eq:NLSE}
    i \frac{\partial \psi}{\partial t} = - \frac{\partial^2 \psi}{\partial x^2} + \kappa |\psi|^2 \psi \ ,
\end{equation}
where $\psi = \psi(x,t)$ is a complex-valued function for all $(x,t) \in [-3\pi,3\pi] \times (0,T=3]$. We set $\kappa = -2$ (focusing case) and seek the bright soliton solution:
\begin{equation}
\begin{split}\label{eq:soliton}
    \psi(x,t) = &q_0\,\text{sech}[q_0(x - x_0 - vt)]\\ &\exp[\frac{i}{2}v(x - x_0) + i t(q_0^2 - \frac{1}{4}v^2)] \ ,
\end{split}
\end{equation}
where $q_0$ and $x_0$ are the initial amplitude and position, and $v$ is the propagation velocity.

Following the procedure in Appendix~\ref{app:space-time}, the discretization of the time and space variables leads to:
\begin{equation}
    \mathbb{O}_x = \mathbb{I}_x - i \tau (\partial^2_{x} + 2 |\psi|^2_x) ,
\end{equation}
such that $\mathbb{O}_x \boldsymbol{\psi}_{j+1} = \boldsymbol{\psi}_{j}$. This gives rise to the following \textit{all-at-once} linear system:
\begin{equation}\label{eq:time_space_linsys}
    (\mathbb{I}_t \otimes \mathbb{O}_x - \mathbb{S}_t \otimes \mathbb{I}_x) \boldsymbol{\psi} = \boldsymbol{b} \ .
\end{equation}
The results obtained using this encoding are reported in Appendix~\ref {app:numerical_results}.

\begin{figure}[t!]
\centering
    \includegraphics[width=.45\linewidth]{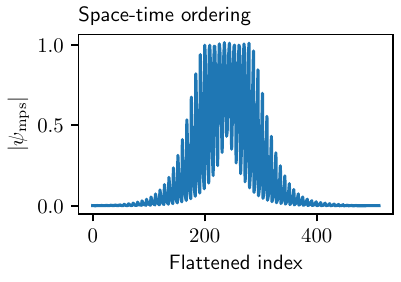}
    \includegraphics[width=.45\linewidth]{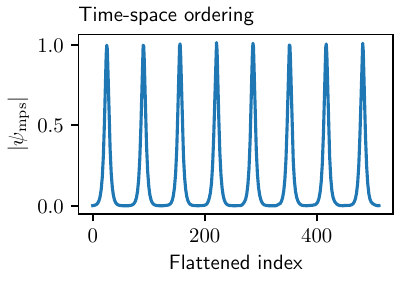}
    \caption{\textbf{Flattened vectors for different MPS encodings.}
    We report the magnitude of the MPS solution for the space-time (left) and the time-space (right) orderings on a reduced domain of $64\times8$ points ($N_x=6$, $N_t=3$). While the space-time solution is oscillatory, the time-space solution results in a smooth function. The grid was chosen for the ease of visualization, highlighting the difference between the encodings.
    }
    \label{fig:time_space_nlse_sol}
\end{figure}

We now comment on the motivation for using the time-space ordering. 
Empirically, for the NLS equation, we notice that the MPS solver converges better in the iterative linear system solver when the space and time legs are swapped, i.e., switching from the space-time to the time-space ordering. Our intuitive understanding of the improved convergence is shown in Fig.~\ref{fig:time_space_nlse_sol}. We plot the flattened vectors of the entire space-time solution resulting from both orderings. We observe that the time-space ordering results in a smooth solution as long as the single-time snapshots are smooth and have periodic boundary conditions. On the other hand, the space-time ordering results in a highly oscillatory function. Since it is well established that smooth functions exhibit low-rank structure, i.e., compressibility~\cite{lindsey2023multiscale,garcia_ripoll2021quantum}, we posit that the time-space ordering is better suited. However, this behavior is to be further investigated in the context of other index orderings and better linear system solvers.

\begin{figure}[t!]
\centering
    \includegraphics[width=0.9\linewidth]{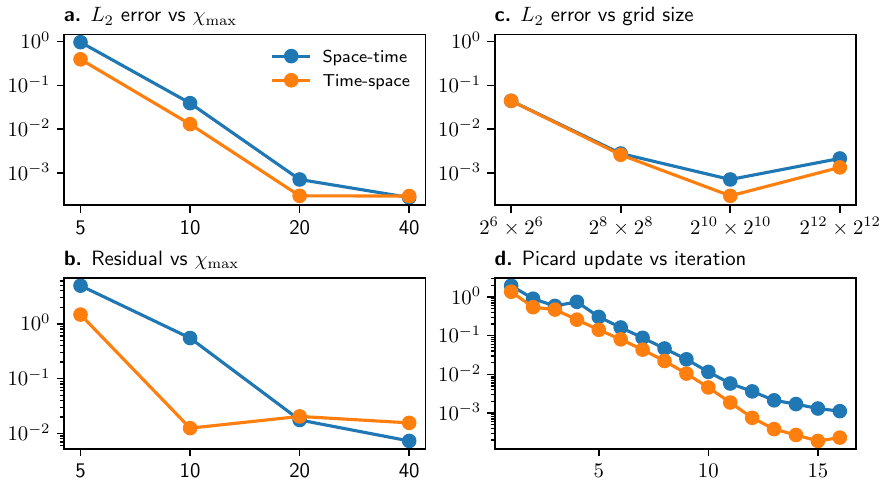}
    \caption{\textbf{Space-time versus time-space ordering for the NLSE solver.} (a)~$L_2$ error and (b)~residual against the bond dimension at $L_x=L_t=10$. (c)~$L_2$ error against the grid size at $\chi_{\max}=20$. (d)~Relative step between Picard iterates at $L_x=L_t=10$ and $\chi_{\max}=20$. All other settings are as in Table~\ref{tab:equations}.}
    \label{fig:nlse_ordering_solver}
\end{figure}

In Fig.~\ref{fig:nlse_ordering_solver}, we measure the effect on the MPS solver by solving the NLSE in both orderings with the same operators, initial condition, Picard loop, and settings ($2^{10}\times2^{10}$ grid, $\chi_{\max}=20$, 16 Picard iterations, 4 DMRG sweeps), varying one parameter at a time. The two runs give the same solution up to the solver's accuracy, and their wall-clock times agree to within 10\%.  The time-space ordering reaches a given accuracy at a smaller bond dimension. The Picard iteration also converges faster in the time-space ordering. On refining the grid at $\chi_{\max}=20$, both orderings have their smallest error at $L_x=L_t=10$ and lose accuracy at $L_x=L_t=12$. The advantage of the time-space ordering is moderate: a factor of $2.4$ in the error at $\chi_{\max}=20$, and none at $\chi_{\max}=40$. Here, we emphasize the fact that this comparison was made only for the NLSE, for a single initial condition and a single run per setting. The same experiment has to be run for all the other equations before drawing a general conclusion.

\section{Extended prediction horizon of MPS-DMD}\label{app:dmd_horizon}
\begin{figure}
    \centering
    \includegraphics[width=0.8\linewidth]{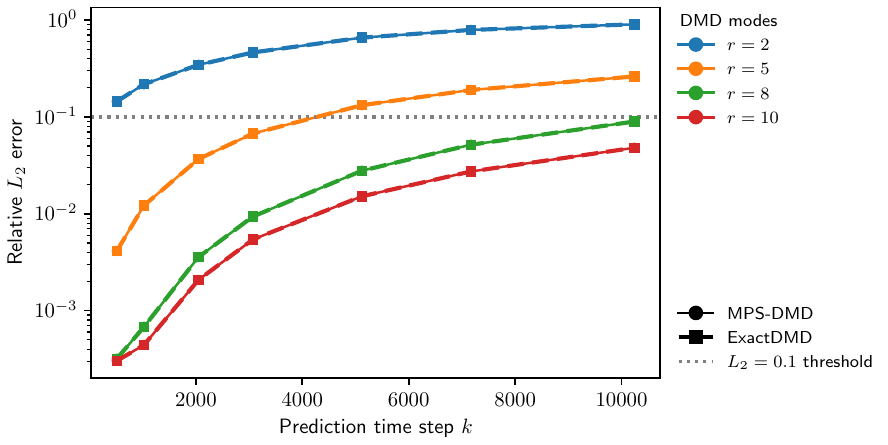}
    \caption{\textbf{Extended prediction horizon of MPS-DMD.} Relative $L_2$ error of the predicted solution of the 1D Burgers' equation against the number of predicted time steps $k$ for $r$ DMD modes, for MPS-DMD (lines) and the exact DMD of PyDMD (dashed). The training window is the 1024 steps of the MPS space-time solution. The dotted line marks an error of $0.1$.}
    \label{fig:dmd_horizon}
\end{figure}

In Fig.~\ref{fig:dmd_horizon}, we extend the MPS-DMD prediction of the 1D Burgers' equation of Fig.~\ref{fig:dmd_prediction} from $k=512$ to $k=10240$ time steps, i.e., ten times the training window of 1024 steps, for $r=2,5,8,10$ modes. The error of MPS-DMD and of the exact DMD of PyDMD agree to within $0.5\%$ at every point shown. The error grows with $k$ for all $r$. 

\end{appendix}

\bibliography{references}

\end{document}